# X-ray computed tomography reconstruction algorithm for refractive index gradient

Keliang Liao[1,#], Qili He[2,#], Panyun Li[1], Liang Luo[3], Peiping Zhu[1,2,*]

[1] Jinan Hanjiang Opto-electronics Technology Company Ltd., Jinan, China
[2] Institute of High Energy Physics, Chinese Academy of Sciences, Beijing, China
[3] School of Aerospace Engineering, Tsinghua University, Beijing, China

[*] Corresponding Author: zhupp@ihep.ac.cn
[#] These authors contributed equally

**Abstract.** The aim of this research is to reconstruct the 3D X-ray refractive index gradient maps by the proposed vector Radon transform and its inverse, assuming that the small-angle deviation condition is met. Theoretical analyses show that the X-ray beam can be modeled as a streamline with continuous change of direction in a row when measured in one grating period, which allows the extraction of the refraction angle signals. Experimental results show that all the 2D refraction signals of different directions can be acquired by a standard circular scanning procedure, which is typically used in the X-ray differential phase-contrast computed tomography. Furthermore, the 3D refractive index gradient maps that contain the directional density changes, can also be accurately reconstructed.

## 1 Introduction

In general, there are two means to reveal the refractive information of samples from the acquired differential phase contrast (DPC) signals. One is to reconstruct the refractive index, and the other is to reconstruct the gradient of the refractive index. Due to its consistency with the classical CT image reconstruction algorithm [1,2], by far, the first approach has been widely discussed [3-12]. Whereas, only few research interests were attracted onto the second approach [13,14]. The main reason is the refractive index gradient reconstruction algorithms lack rigorous theoretical foundation for the X-ray DPC imaging. To overcome this difficulty, a new vector Radon and inverse Radon analysis theory has been proposed for the first time in this paper, which takes into account the X-ray beam small angle deviation condition. Based on this new proposed theory, the analytical algorithm for 3D refractive index gradient has been derived strictly.

## 2  Model and methodology

The flow chart of the proposed model in this paper can be summarized in Fig.1. The small angle deviation condition is a crucial process to establish this new theory. In the following section, the detail of each process is presented.

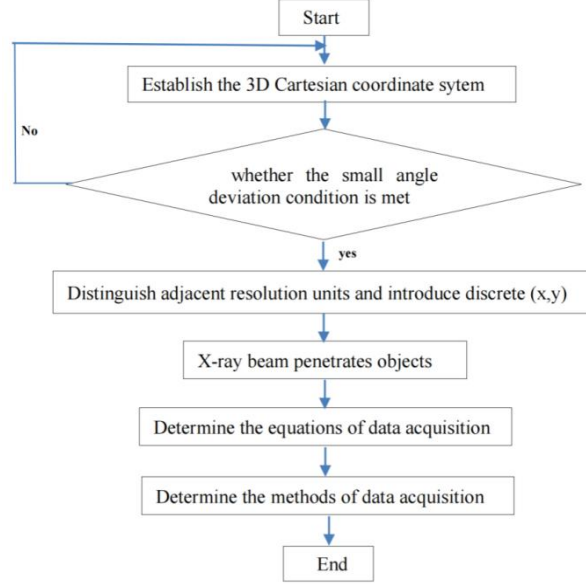

Fig.1. The flow chart of the proposed model in this study.

### 2.1  Optical streamline model

According to the Maxwell's electromagnetic equations, the Helmholtz equation for scalar monochromatic wave in inhomogeneous medium can be derived. Furthermore, both the Eikonal equation and the continuity equation [15] can be obtained. Assuming an infinitely small wavelength, the differential equation of light rays in geometric optics [15,16] can be expressed as follows:

$$\frac{d}{ds}[\nabla S(\vec{r})] = \frac{d}{ds}[n(\vec{r})\vec{\tau}(\vec{r})] = \nabla n(\vec{r}), \tag{1}$$

where $s$ is the arc length of the beam, $S$ is the optical path length, $\vec{r}$ is the position vector, $\vec{\tau} = d\vec{r}/ds$ is the unit vector parallel to the tangent of the beam, and $n\vec{\tau}$ is the ray vector, and $n$ is the refractive index of the medium

$$n(\vec{r}) = 1 - \delta(\vec{r}) + i\beta(\vec{r}), \tag{2}$$

where $\delta$ denotes the decrement of the real portion and is related to the beam phase shift, $\beta$ denotes the imaginary portion and is related to the beam attenuation. Substituting Eq. (2) into Eq. (1), the following results can be obtained:

$$\frac{d}{ds}[\nabla S(\vec{r})] = \frac{d}{ds}\vec{\tau}(\vec{r}) = -\nabla \delta(\vec{r}). \tag{3}$$

In Eq. (3), both the δ and β are ignored for the middle term, and β is ignored for the last term due to the fact that $\beta \ll \delta \ll 1$, and the ray vector $n\vec{\tau}$ reduced to $\vec{\tau}$ whose increment is in effect the increment of the refraction angle of the X-ray beam inside the sample, as illustrated in Fig.2.

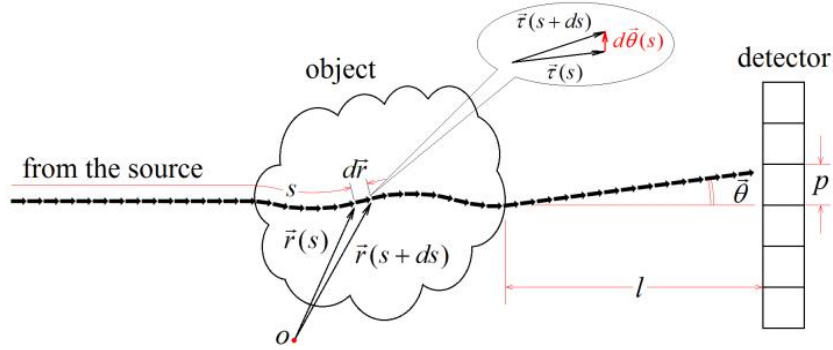

Fig.2. Illustration of the X-ray beam passing through the object when considering the internal small refractions. The $l$ denotes the distance between the sample and the detector, and $p$ denotes the dimension of the detector element.

Assuming that a group of parallel X-ray beam penetrates through the object with a refractive index n, shown in Fig.2. Based on Eq.(3), the integration of the refraction index gradient $\nabla \delta$ along a real X-ray beam path (not straight line any longer) is expressed as

$$\nabla S(s) - \nabla S(-\infty) = \vec{\theta}(s) = -\int_{-\infty}^{s} \nabla \delta(\vec{r}) ds. \quad (4)$$

The relation between phase and optical path

$$\Phi(s) = \frac{2\pi}{\lambda} S(s) = k S(s) \quad (5)$$

is substituted into Eq.(4), one gets

$$\frac{\nabla \Phi(s) - \nabla \Phi(-\infty)}{-k} = -\vec{\theta}(s) = \int_{-\infty}^{s} \nabla \delta(\vec{r}) ds. \quad (6)$$

Alternatively, Eq.(6) can also be rewritten as

$$\frac{\nabla \Phi[s(x,y,z)] - \nabla \Phi(-\infty)}{-k} = -\vec{\theta}[s(x,y,z)] = \int_{-\infty}^{s(x,y,z)} \nabla \delta(x,y,z) ds. \quad (7)$$

Herein, the standard Cartesian coordinate system is selected with the z axis parallel to the incident direction. Due to the inhomogeneous object refractive index distribution, as illustrated in Fig.2, the exit X-ray beam would be deviated by $\vec{\theta}$ compared to its primary incident direction. However, in the process of experimental data analysis, it is quite difficult to measure the non-zero angular deviations induced by inhomogeneous object. The reason is that the pixel size of the detector is limited. For instance, if $|\vec{\theta}| < p/l$, the detection system with detector element size of $p$ would be insensitive in probing such tiny beam deviations. In other words, it is difficult for the detector element with finite size to distinguish the refracted X-ray beams from the primary X-ray beams when they are falling within the same resolution unit (suppose the resolution unit size is 2p). When this happens, essentially, the change of X-ray beam inside the x-y plane should be neglected, the arc length of the beam $s(x, y, z)$ can be reduced to $z$. In order to distinguish the adjacent resolution units, discrete imaging unit $(x, y)$ according to resolution elements are introduced into the functions of phase gradient and refraction angle, thus equation (7) can be simplified as

$$\frac{\nabla\Phi(x,y,z)-\nabla\Phi(-\infty)}{-k} = -\vec{\theta}(x,y,z) = \int_{-\infty}^{z}\nabla\delta(x,y,z)dz. \quad (8)$$

Please note that since the integration of $\nabla\delta$ is the path integral in front of the detector, $(x, y)$ inside the integral will not be affected by the detector and is still continuous. Therefore, continuity and discreteness coexist in Eq.(8), inside and outside the integral respectively. Once the X-ray beam leaves the object, then $\nabla\delta = 0$. Thus, by letting $z \to \infty$, one gets

$$\vec{e}_z \frac{\partial \Phi(x,y,\infty)}{\partial z} - \vec{e}_z \frac{\partial \Phi(-\infty)}{\partial z} = 0, \quad (9)$$

and takes account of the fact

$$\vec{e}_x \frac{\partial \Phi(-\infty)}{\partial x} + \vec{e}_y \frac{\partial \Phi(-\infty)}{\partial y} = 0, \quad (10)$$

where $\vec{e}_x$, $\vec{e}_y$ and $\vec{e}_z$ are the unit vector along the $x$, $y$ and $z$ axis respectively, Eq.(8) can be reduced as

$$\frac{\nabla_\perp \Phi(x,y)}{-k} = -\vec{\theta}(x,y) = \int_{-\infty}^{\infty}\nabla\delta(x,y,z)dz, \quad (11)$$

where $\nabla_\perp = \vec{e}_x \partial/\partial x + \vec{e}_y \partial/\partial y$. Note that $|\vec{\theta}(x,y)| < p/l$ is defined as the small angle deviation condition in this work, which solves the contradiction between straight line propagation required by Radon transform and directional change propagation caused by refraction angle. Under this condition, the refraction angle in Eq.(11) is a vector sum (integral) process, as shown in Fig.3.

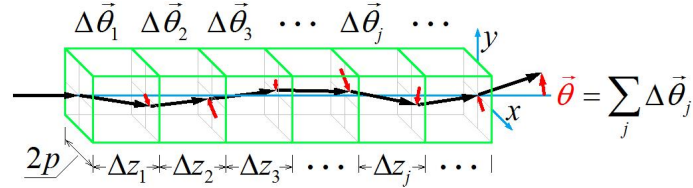

Fig.3. Diagram of refraction angle formation.

## 2.2 Vector reconstruction algorithm for refractive index gradient

Assuming a parallel X-ray CT imaging geometry with the $y$ axis being the axis of sample rotation, the analytical refractive index gradient $\nabla\delta$ reconstruction algorithm is discussed. Let $y' = y$, and $(x', y', z')$ be the object coordinate, as a result, Eq.(11) can be expressed as the Radon transformation of $\nabla\delta$ at the view angle $\varphi$ between $x'$ axis and $x$ axis, namely

$$\frac{\nabla_\perp \Phi(x,y,\varphi)}{-k} = -\vec{\theta}(x,y,\varphi) = \int_{-\infty}^{\infty}\int_{-\infty}^{\infty}\nabla\delta(x',y',z')\,\hat{\delta}(x'\cos\varphi + z'\sin\varphi - x)dx'dz', (12)$$

where $\hat{\delta}$ is the Dirac pulse function. Afterwards, applying the inverse Radon operation onto Eq.(12), the $\nabla\delta$ can be readily reconstructed,

$$\nabla\delta(x',y',z') = \vec{e}_{x'}\frac{\partial\delta(x',y',z')}{\partial x'} + \vec{e}_{y'}\frac{\partial\delta(x',y',z')}{\partial y'} + \vec{e}_{z'}\frac{\partial\delta(x',y',z')}{\partial z'} = -$$
$$\int_0^\pi d\varphi \int_{-\infty}^{\infty} \mathcal{F}_x^{-1}|\rho|\mathcal{F}_x[\theta_x(x,y,\varphi)\vec{e}_x + \theta_y(x,y,\varphi)\vec{e}_y]\hat{\delta}(x'\cos\varphi + z'\sin\varphi - x)dx, \quad (13)$$

where $\vec{e}_{x'}$, $\vec{e}_{y'}$ and $\vec{e}_{z'}$ are the unit vector along the $x'$, $y'$ and $z'$ axis, respectively; $\theta_x = \partial\Phi/k\partial x$ and $\theta_y = \partial\Phi/k\partial y$ are the component of the refraction angle on the $x$ and $y$ axis, respectively, the operators $\mathcal{F}_x$ and $\mathcal{F}_x^{-1}$ represent the Fourier and inverse Fourier transformation operator along $x$ axis, correspondingly; and variable $\rho$ denotes the frequency counterpart of space variable $x$.

### 2.3 DPC imaging with inclined phase and analyzer gratings

Despite the explicit reconstruction expression of $\nabla\delta$ shown in Eq.(13), it is still very challenging to apply it on real experimental data analysis. This is because the refraction angle signals along the x axis and y axis are both required to apply into Eq.(13). In practice, the easiest way to obtain the two perpendicular components of refraction angle is to rotate the grating interferometry with respect to the $z$ axis by 90 degrees. Obviously, this may bring inconvenience to the data acquisition and does not meet the fast imaging demand. To overcome such difficulty, we proposed one alternative method to acquire the $\theta_x\vec{e}_x + \theta_y\vec{e}_y$ data with the grating interferometry inclined by $\omega$ degrees. In this method, $\omega$ is the angle between $x''$ axis and $x$ axis, and $x''$ axis is perpendicular to the 1D grating groove, see the proposed grating settings in Fig.4.

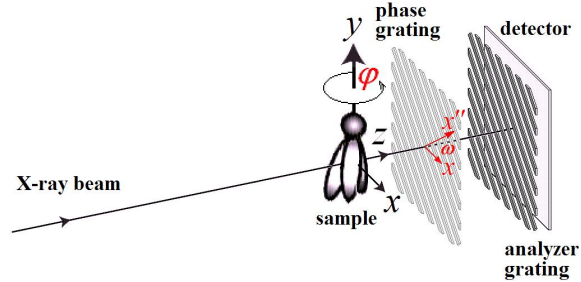

Fig.4. Illustration of the DPC imaging with inclined phase and analyzer gratings. The variable $\omega$ represents the angle between $x''$ axis and x axis and the $x''$ axis is parallel to the normal vector of the 1D grating bar.

With such special grating alignments, the refraction angle signal of object acquired at $\omega$ is equal to $\theta_{x''}(x,y,\varphi)$ see Fig.5 (a). If rotating the object with respect to the y axis by 180 degrees, as shown in Fig.5 (b), the measured refraction angle signal becomes $\theta_{x''}(-x,y,\varphi+\pi)$, which is equal to $\theta_{y''}(x,y,\varphi)$, as shown in Fig.5(c). As a result, the needed bilateral refraction angle signals along two perpendicular directions can be acquired with continuous object rotation. In all, we have

$$\theta_x(x,y,\varphi) = \frac{\theta_{x''}(x,y,\varphi) - \theta_{y''}(x,y,\varphi)}{2\cos\omega} = \frac{\theta_{x''}(x,y,\varphi) - \theta_{x''}(-x,y,\varphi+\pi)}{2\cos\omega}, \quad (14)$$

$$\theta_y(x, y, \varphi) = \frac{\theta_{x''}(x,y,\varphi)+\theta_{y''}(x,y,\varphi)}{2\sin\omega} = \frac{\theta_{x''}(x,y,\varphi)+\theta_{x''}(-x,y,\varphi+\pi)}{2\sin\omega}. \quad (15)$$

By substituting the Eq.(14) and Eq.(15) back into the Eq.(13), one obtains

$$\nabla\delta(x', y', z') = \vec{e}_{x'}\frac{\partial\delta(x',y',z')}{\partial x'} + \vec{e}_{y'}\frac{\partial\delta(x',y',z')}{\partial y'} + \vec{e}_{z'}\frac{\partial\delta(x',y',z')}{\partial z'} = -$$
$$\int_0^\pi d\varphi \int_{-\infty}^\infty \mathcal{F}_x^{-1} |\rho| \mathcal{F}_x \left[ \frac{\theta_{x''}(x,y,\varphi)-\theta_{x''}(-x,y,\varphi+\pi)}{2\cos\omega}\vec{e}_x + \frac{\theta_{x''}(x,y,\varphi)+\theta_{x''}(-x,y,\varphi+\pi)}{2\sin\omega}\vec{e}_y \right] \hat{\delta}(x'\cos\varphi + z'\sin\varphi - x)dx. \quad (16)$$

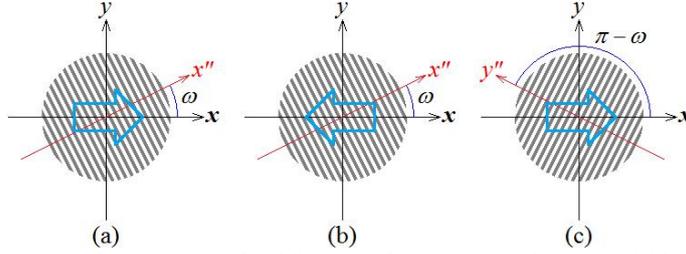

(a)      (b)      (c)

Fig.5. Scheme of the proposed fast bilateral refraction angle signal acquisition method with the positions of both grating and object: (a) $\omega$, $\varphi$; (b) $\omega$, $\varphi + \pi$; (c) $\pi - \omega$, $\varphi$. The blue arrow indicates the object.

Assuming a unit vector in the object space
$$\vec{e} = \vec{e}_{x'}\sin\gamma\cos\psi + \vec{e}_{y'}\cos\gamma + \vec{e}_{z'}\sin\gamma\sin\psi, \quad (17)$$
where $\gamma$ is the latitude, $0 \le \gamma < \pi$, and $\psi$ is the longitude, $0 \le \psi < 2\pi$. Readily, the 1D projection gradient of $\nabla\delta$ along the $\vec{e}$ vector is derived to be equal to
$$\vec{e}[\vec{e} \cdot \nabla\delta(x', y', z')] = \vec{e}\left[ \sin\gamma\cos\psi\frac{\partial\delta(x',y',z')}{\partial x'} + \cos\gamma\frac{\partial\delta(x',y',z')}{\partial y'} + \sin\gamma\sin\psi\frac{\partial\delta(x',y',z')}{\partial z'} \right]. \quad (18)$$

Moreover, letting the vector $\vec{e}$ be parallel with the normal of the selected plane, then the 2D projection gradient of $\nabla\delta$ onto this plane is
$$\nabla\delta(x', y', z') - \vec{e}[\vec{e} \cdot \nabla\delta(x', y', z')] = \vec{e}_1[\vec{e}_1 \cdot \nabla\delta(x', y', z')] + \vec{e}_2[\vec{e}_2 \cdot \nabla\delta(x', y', z')] =$$
$$\vec{e}_1\left[-\cos\gamma\cos\psi\frac{\partial\delta(x',y',z')}{\partial x'} + \sin\gamma\frac{\partial\delta(x',y',z')}{\partial y'} - \cos\gamma\sin\psi\frac{\partial\delta(x',y',z')}{\partial z'}\right] +$$
$$\vec{e}_2\left[-\sin\psi\frac{\partial\delta(x',y',z')}{\partial x'} + \cos\psi\frac{\partial\delta(x',y',z')}{\partial z'}\right], \quad (19)$$

where $\vec{e} \perp \vec{e}_1$,
$$\vec{e}_1 = -\vec{e}_{x'}\cos\gamma\cos\psi + \vec{e}_{y'}\sin\gamma - \vec{e}_{z'}\cos\gamma\sin\psi, \quad (20)$$
and
$$\vec{e}_2 = \vec{e} \times \vec{e}_1 = -\vec{e}_{x'}\sin\psi + \vec{e}_{z'}\cos\psi. \quad (21)$$

## 3 Experimental validation

### 3.1 Experimental setup

Validation experiments were performed on the Talbot interferometer system of the BL13W1 beamline at Shanghai Synchrotron Radiation Facility (SSRF). As shown in Fig.6, it consists of one 1D phase grating ($0.5\pi$ shifting, period 2.396mm) and one 1D absorption grating (period 2.400mm) . The distance between the two gratings is 46.380 mm. The beam energy is 20 keV. The detector pixel size is 6.5mm. The specimen of a hamster front toe was positioned with its rotation axis along $y$ axis with the grating interferometry inclined by $\omega = 45°$. For this Talbot interferometer on SSRF, the average fringe visibility is about 0.40, and the mean detector readout is around 25,000. The projection images of object were acquired from projection angle 0 degree to 360 degrees with 0.5 degree interval. At each projection angle the phase-stepping scan was acquired by translating the absorption grating along the $x''$ axis with a exposure time of 13 ms in 8 equidistant steps over one grating period. The bilateral refraction angle signals were extracted from two angular intervals: $\theta_{x''}(x,y,\varphi)$ for $\varphi$ ranges from 0 to $\pi$, and $\theta_{x''}(-x,y,\varphi)$ for $\varphi$ ranges from $\pi$ to $2\pi$. The programming language used for reconstruction is MATLAB. It takes about 62 seconds to reconstruct a transverse map, and the COMPUTER CPU is Intel Xeon E5-2667*2.

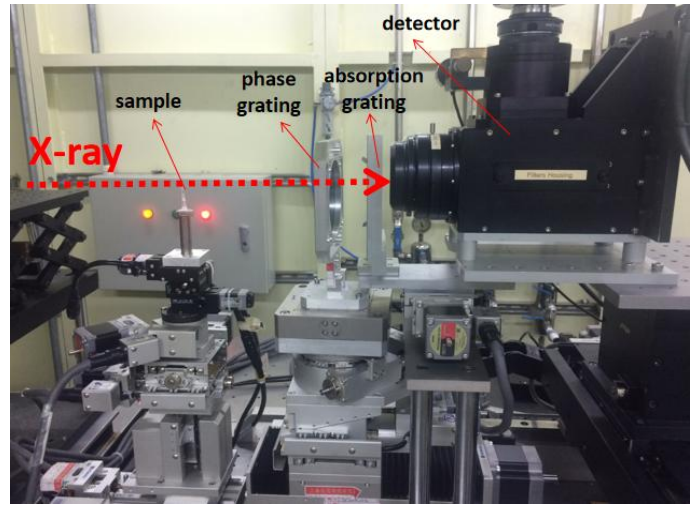

Fig.6. The experimental setup of the Talbot interferometer system.

### 3.2 Results

Experimental imaging result of $|\nabla\delta|$ is shown in Fig.7. The three-dimensional distribution of $|\nabla\delta|$ is presented in the $x'$-$y'$-$z'$ Cartesian coordinate system, with $x'$-$z'$

plane as the transverse plane, *x'-y'* plane as the coronal plane, and *y'-z'* plane as the sagittal plane. From this scalar map of $|\nabla\delta|$, the details of the structural information can be obtained, but the directional information about the sample is lost. To reveal the directional information of object, the vector information of $\nabla\delta$ will be discussed in detail in this following section.

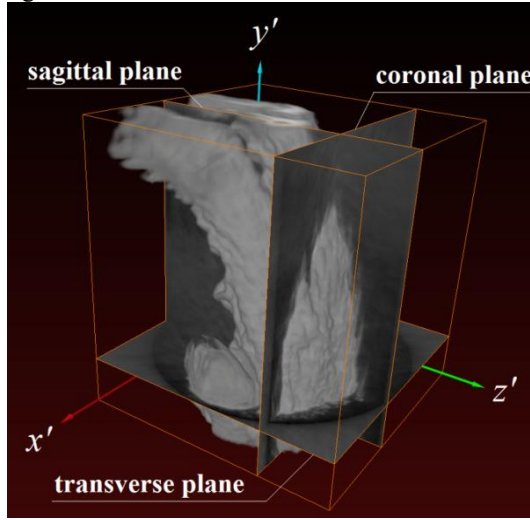

Fig.7. Experimental imaging result of $|\nabla\delta|$.

In particular, Fig.8 illustrates two special 1D projection gradients of $\nabla\delta$ in the coronal plane: the $\vec{e}_{y'} = \partial\delta(x',y',z')/\partial y'$ is parallel to the hair, and the $\vec{e}_{x'} = \partial\delta(x',y',z')/\partial x'$ is perpendicular to the hair. Clearly, the hairs can be easily identified from Fig.8(b), while hair is almost invisible in Fig.8(a), which shows the important ability of the proposed vector CT algorithm when compared with the traditional scalar CT algorithms.

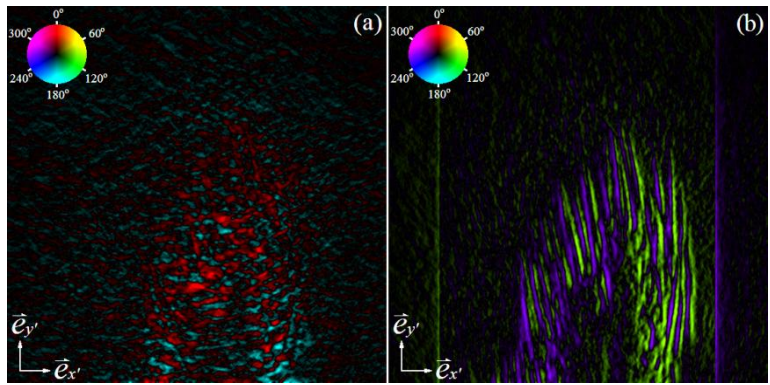

Fig.8. The image of 1D projection gradient of $\nabla\delta$ on the coronal plane with parallel (a) and perpendicular (b) component, correspondingly.

In addition, the 2D projection gradient maps of $\nabla\delta$ on the coronal, sagittal, and transverse planes are illustrated in Fig.9(a) to Fig.9(c), respectively. In these reconstructed projection maps, the color represents the signal orientation. The gradient variations of the refractive index can be clearly distinguished from the different color and brightness distributions. Fig.9(a) shows the 2D projection gradient in the coronal plane, where $\vec{e} = -\vec{e}_{z'}$, $\vec{e}_1 = \vec{e}_{y'}$, $\vec{e}_2 = \vec{e}_{x'}$. Fig.9(b) shows the 2D projection gradient in the sagittal plane, where $\vec{e} = \vec{e}_{x'}$, $\vec{e}_1 = \vec{e}_{y'}$, $\vec{e}_2 = \vec{e}_{z'}$. Fig.9(c) shows the 2D projection gradient in the transverse plane, where $\vec{e} = \vec{e}_{y'}$, $\vec{e}_1 = \vec{e}_{z'}$, $\vec{e}_2 = \vec{e}_{x'}$. Moreover, the hue saturation value (HSV) color map is used to depict such projection gradient information, including both the absolute signal strength and its angular orientation, into one single image [17]. More specifically, the hue corresponds to the angle; the saturation is set to 1 and the value is defined as the normalized brightness of the directionality in order to fill the span [0, 1], meaning that dark areas in the image correspond to areas with no gradient. For instance, the red, yellow, green, cyan, blue, magenta colors correspond to the 0°, 60°, 120°, 180°, 240°, 300° angular direction, respectively.

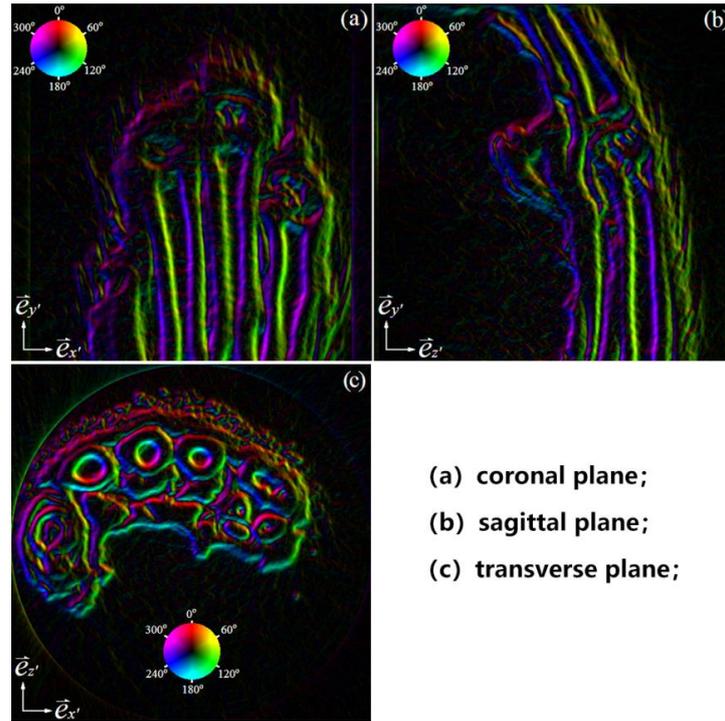

(a) coronal plane;
(b) sagittal plane;
(c) transverse plane;

Fig.9. The 2D projection gradient in the coronal plane (a), the sagittal plane (b) and the axial plane (c).

As a consequence, the line integral process of $\nabla\delta$ in the Radon transform, equals to the 2D vector summation of a collection of the increment of refraction angles along the real X-ray beam propagation path, see Fig.3 and Eq.(11). However the principle of the inverse Radon transform of $\vec{\theta}$ is to filter these 2D vector summation of each projection direction and back-project them to reconstruct the 3D refractive index gradient $\nabla\delta$, see Eq.(13) or (16). As shown in Fig.7 to Fig.9, both the Radon and inverse Radon transforms can be performed on vectors in refraction angle signal based computed vector tomography.

### 3.3 Discussions

Based on the small angle deviation condition, the Radon transform essentially depends on the pixel size of detector. When the pixel element size reduces, the small angle deviation condition approximates to the rigorous straight line propagation condition. In other words, the ideal straight-line propagation condition is the theoretical limit of the small angle deviation condition. Therefore, the deflected X-ray beam induced by the density changes of object is made of two parts: the straight-line component and the streamline component. Obviously, the straight-line component of X-ray beam, which is measured within one resolution unit, meets the requirements of Radon transform. While the streamline component with continuous change of direction, which is measured by the grating period, meets the requirements of extracting the refraction angle signals.

## 4 Conclusion and Outlook

In this study, the small angle condition was proposed to define the Radon transform of the refractive index gradient. Based on this condition, analytical reconstruction algorithm for the refractive index gradient and the corresponding data acquisition method were developed. In the verification experiment, complete projection data of refractive index gradient were collected by using grating interferometer with inclined degree $\omega = 45°$. Further, the 3D refractive index gradient was reconstructed to verify this new CT image reconstruction theory. Experimental results show two major advantages of this proposed CT theory. First, a standard circular acquisition trajectory typically used in conventional X-ray computed tomography can measure bilateral refraction angle signals. Second, the reconstructed 3D refractive index gradient maps enhance the visualization of the direction of density changes with the flexibility.

Above all, the proposed new reconstruction algorithm has huge promising application scenarios, for example in materials testing of fibrous composites and in medical diagnosis. Besides, the theory and method established in this work can also be applied for other imaging fields, like the neutron imaging, proton imaging, electron imaging, optical imaging, and so on.

**Acknowledgments.** This work is supported by the Jinan Haiyou Industry Leading Talent Project(2022), the Innovation Promotion Project of SME in Shandong Province(No.2023TSGC0093), the Enterprise Technology Innovation Project of Shandong Province(No.202350100372), the National Natural Science Foundation of China (Grant No. 11535015).